\documentclass[aps,showpacs,prl,twocolumn,superscriptaddress,monochrome]{revtex4-1}
\usepackage{graphicx}

\usepackage{color}
\usepackage{siunitx}
\usepackage{dsfont}
\usepackage{chemformula}
\usepackage{hyperref}
\graphicspath{{images/}}
\makeatletter
\def\maketitle{
\@author@finish
\title@column\titleblock@produce
\suppressfloats[t]}
\makeatother
\begin{document}

\title{
Reconciling the theoretical and experimental electronic structure of NbO$_2$
}

\author{Samuel Berman}
\email[]{bermans@tcd.ie}
\affiliation{School of Physics and Centre for Research on Adaptive Nanostructures and Nanodevices (CRANN), Trinity College Dublin, The University of Dublin, Dublin 2, D02 PN40 Ireland}
\author{Ainur Zhussupbekova}
\affiliation{School of Physics and Centre for Research on Adaptive Nanostructures and Nanodevices (CRANN), Trinity College Dublin, The University of Dublin, Dublin 2, D02 PN40 Ireland}
\affiliation{School of Chemistry, Trinity College Dublin, The University of Dublin, Dublin 2, D02 PN40 Ireland}
\author{Jos E. Boschker}
\affiliation{Leibniz-Institut für Kristallzüchtung, Max-Born-Str. 2, 12489 Berlin, Germany}
\author{Jutta Schwarzkopf}
\affiliation{Leibniz-Institut für Kristallzüchtung, Max-Born-Str. 2, 12489 Berlin, Germany}
\author{David D. O’Regan}
\affiliation{School of Physics and Centre for Research on Adaptive Nanostructures and Nanodevices (CRANN), Trinity College Dublin, The University of Dublin, Dublin 2, D02 PN40 Ireland}
\author{Igor V. Shvets}
\affiliation{School of Physics and Centre for Research on Adaptive Nanostructures and Nanodevices (CRANN), Trinity College Dublin, The University of Dublin, Dublin 2, D02 PN40 Ireland}
\author{Kuanysh Zhussupbekov}
\email[]{zhussupk@tcd.ie}
\affiliation{School of Physics and Centre for Research on Adaptive Nanostructures and Nanodevices (CRANN), Trinity College Dublin, The University of Dublin, Dublin 2, D02 PN40 Ireland}
\affiliation{School of Chemistry, Trinity College Dublin, The University of Dublin, Dublin 2, D02 PN40 Ireland}
\affiliation{Kh. Dosmukhamedov Atyrau University, Studenchesky Ave.,1, Atyrau, Kazakhstan}

\begin{abstract}
\textcolor{red}{Metal-insulator transition materials such as NbO$_2$ have generated much excitement in recent years for their potential applications in computing and sensing. NbO$_2$ has generated considerable debate over the nature of the phase transition, and the values for the band gap/band widths in the insulating phase.} We present a combined theoretical and experimental study of the band gap and electronic structure of the \textcolor{red}{insulating} phase of NbO$_2$. We carry out \textit{ab-initio} density functional theory plus $U$ calculations, directly determining $U$ and $J$ parameters for both the Nb 4\textit{d} and O 2\textit{p} subspaces through the recently introduced minimum-tracking linear response method. We find a fundamental bulk band gap of 0.80 eV for the full DFT+$U$+$J$ theory. We also perform calculations and measurements for a (100) oriented thin film. Scanning tunnelling spectroscopy measurements show that the surface band gap varies from 0.75 eV to 1.35 eV due to an excess of oxygen in and near the surface region of the film. \textcolor{red}{Slab calculations indicate metallicity localised at the surface region caused by an energy level shift consistent with a reduction in Coulomb repulsion.} We demonstrate that this effect in combination with the simple, low cost DFT+$U$+$J$ method can account for the band widths and \textit{p}-\textit{d} gap observed in X-ray photoelectron spectroscopy experiments. \textcolor{red}{Overall, our results indicate the possible presence of a 2D anisotropic metallic layer at the (100) surface of NbO$_2$.}

\end{abstract}


\maketitle
\section{I. Introduction}

Materials that undergo a metal to insulator transition (MIT) attract considerable attention due to their potential applications in areas such as memory devices \cite{Kim2013,Cha2016}, sensing \cite{Strelcov2009,Chen2019}, and neural-inspired computing \cite{Kumar2017,Pickett2013}. Niobium dioxide (NbO$_2$) undergoes such a transition at 1080 K, with a change in crystal structure from the \textcolor{red}{insulating} distorted rutile, body centred tetragonal (BCT) phase to the metallic rutile phase. Reconciling theoretical calculations with the experimental measurements for this material has proven challenging, most notably in regard to the fundamental (indirect) band gap of the \textcolor{red}{insulating} BCT phase. Theoretical calculations of the band gap range from 0.058 eV \cite{Wahila2019} to 1.48 eV \cite{OHara2014}. \textcolor{red}{Experimental measurements also show a broad range}, from 0.7 eV \cite{OHara2014} to 1.23 eV \cite{Sakai1985}. Similar discrepancies between theory and experiment exist for other spectral properties, such as the widths of the Nb 4\textit{d$_{xy}$} and O 2\textit{p} bands, as well as the \textit{p}-\textit{d} gap. The present work seeks to resolve this discrepancy. 

Experimentally, the large variation in band gap can be due to many factors, such as crystalline quality and the technique used to measure the gap. The smallest gaps are reported for room temperature optical measurements, usually around 0.7 eV \cite{OHara2014}. In this case the temperature will certainly effect the measured gap, as well as exciton binding effects. The highest reported gap was measured on a polycrystalline sample, where a value for the gap of 1.23 eV was extracted from resistance versus temperature measurements \cite{Sakai1985}. The sample used in this measurement was ceramic NbO$_2$ obtained by reducing an Nb$_2$O$_5$ sample under a hydrogen atmosphere. It is therefore highly likely that residual Nb$_2$O$_5$ in the sample would effect the measured gap. Combined photoemission and inverse photoemission spectroscopy (PES and IPES) experiments determined a lower bound on the band gap to be 1.0 eV \cite{Posadas2014}. However, core level XPS measurements also showed the presence of two oxidation states of Nb on the surface, indicating a possible capping layer. Due to the extreme surface sensitivity of IPES this could significantly impact the measured band gap. The small thickness of the film used (4 nm) could also give rise to quantum size effects, distorting the true gap of the bulk phase. For these reasons, it is uncertain whether this PES/IPES measurement can be taken as a true representation of the bulk NbO$_2$ band structure. Recently, Stoever et al. \cite{Stoever2020} finally obtained strong agreement between the optical gap measured at low temperature (0.85 eV) and the electronic gap determined from resistance versus temperature measurements (0.88 eV), for high quality bulk-like NbO$_2$ thin films (thickness of 100 nm). They also obtained a room temperature optical band gap of 0.76 eV, in line with other reports \cite{OHara2014}. Therefore, it would seem that experimentally there is still some degree of uncertainty, but the most reliable measurements for the bulk band gap indicate a value of around 0.85 to 0.88 eV.

With regard to theoretical calculations, the importance of electronic exchange and correlation in NbO$_2$ has become a point of some contention in the literature. For the rutile phase, Brito et al. \cite{Brito2017} performed cluster dynamical mean field theory (cDMFT) calculations \textcolor{red}{(with assumed Nb 4\textit{d} subspace parameters of $U=6$ eV, $J=1$ eV)} and determined that correlation plays a significant role in that phase. For the BCT phase, O'Hara et al. \cite{OHara2014} performed DFT+$U$ calculations varying $U$ from 0 to 5 eV (only on the Nb 4\textit{d} subspace). They found a band gap of 0.83 eV with $U=2$ eV, in line with the lower end of the reported experimental gaps, however the band widths and \textit{p}-\textit{d} gap showed poor agreement with experiment. With an emphasis more on exchange, calculations with the hybrid Heyd–Scuseria–Ernzerhof (HSE) functional from both O'Hara et al. \cite{OHara2014} and Posades et al. \cite{Posadas2014} show a band gap 1.48 eV, much higher than any reported experimental value. Other authors suggest that correlations play at most a minor role in NbO$_2$ \cite{Eyert2002,Kulmus2021}. Kulmus et al. \cite{Kulmus2021} carried out GW plus Bethe-Salpeter (GW-BSE) calculations and found a band gap of 0.98 eV, a slight overestimation on the likely experimental value, yet much improved compared to LDA or HSE. Additionally they found excellent agreement with experiment for their calculated dielectric function. However, in addition to the band gap, the band widths and \textit{p}-\textit{d} gap remain problematic, being underestimated and overestimated, respectively.

In this paper we show that first principles DFT+$U$+$J$ calculations show solid agreement with experiment in terms of the bulk band gap, and through scanning tunnelling spectroscopy (STS) we reveal that excess of oxygen can significantly influence the local surface band gap. Additionally we demonstrate that slab calculations including DFT$+U$+$J$ (which are feasible due to the relatively low computational expense of this method) can predict the correct band widths and \textit{p}-\textit{d} gap observed in X-ray photoelectron spectroscopy (XPS). Overall, this indicates that a relatively simple treatment of exchange and correlation is sufficient for NbO$_2$, when relevant surface effects are included.

\section{II. Experimental and Computational details}

\subsection{A. Experimental Details}

Experimental measurements were performed across several ultra-high vacuum (UHV) systems. The NbO$_{2}$ film was deposited on top of a rutile (110) TiO$_{2}$ substrate via pulsed laser deposition (PLD) of a NbO$_{2}$ target, the purity of which was 99.9985$\%$. An aperture was employed to the central part of a KrF laser (wavelength 248 nm), which was resulting in a power of 65 mJ/pulse and a fluence of 2.1 cm$^{2}$. For the synthesis, a pulse repetition rate of 5 Hz and in total 12000 pulses was used. The distance between the target and substrate was 60 mm and the substrate temperature was 650$^{\circ}$C. A partial pressure of 0.1 mbar of Ar gas was utilised during the synthesis. 

To provide compositional and structural information of the thin NbO$_{2}$ film, XPS, X-ray diffraction (XRD), reflection high-energy electron diffraction (RHEED) and low-energy electron diffraction (LEED) were carried out. RHEED was conducted \textit{in situ} during and after the film  growth. XRD was utilised to analyse the structure and orientation of the NbO$_{2}$ crystal lattice (XRD data is shown in the SI Fig. 1(a,b) \cite{SI1}). The thickness of the NbO$_{2}$ film was estimated at approximately 80 nm by the evaluation of the thickness fringes around the film peak in the $\theta$/2$\theta$ XRD scans. For the XRD measurements, a Bruker D8 HRXRD was used with a Cu anode X-ray tube with a solid-state 1D detector, a G\"{o}bel mirror for converting the diverging output of the tube into a parallel beam, and a two bounce Ge monochromator for the selection the Cu-K$_{\alpha}$1 emission with wavelength of 1.5406 \AA\, from the spectrum. An Omicron MultiProbe XPS system with a monochromatized Al K$_{\alpha}$ source (XM 1000, 1486.7 eV) was used for compositional analysis and resolving a valence band (VB) structure of NbO$_{2}$ film. The base pressure of XPS is 2$\times$10$^{-10}$ mbar, and the instrumental resolution is 0.6 eV. Peak positions and elemental components were selected from the elemental library \cite{Fairley2021} with Gaussian/Lorentzian line shapes and Shirley background. LEED measurements were performed, in conjunction with scanning tunneling microscopy and spectroscopy (STM/S), to analyse the surface structure of the NbO$_{2}$ film grown on top of rutile TiO$_{2}$(110) substrate.

The scanning tunnelling microscope (STM) utilised in this study is a commercial low-temperature system from Createc with a base pressure of $5\times10^{-11}$\,mbar. All STM images were obtained at liquid nitrogen temperature (77\,K) in constant-current mode (CCM). For STS measurements, the tip was moved across the grid with the tip height stabilised by the CCM scanning parameters. Before the spectra was obtained at any point, the feedback loop was turned off such that the current was allowed to be recorded as a function of the voltage, while maintaining a constant tip height. The feedback was subsequently turned back on and the tip was moved to the next grid position \cite{Zhussupbekov2020-1, Zhussupbekov2021-1,Berman2023}. The preparation chamber of the UHV system is fitted with a cooling/heating stage, LEED and an ion gun for sputtering. The STM tips used were of [001]-oriented single-crystalline tungsten, which electrochemically etched in NaOH. The bias was applied to the sample with respect to the tip.

The NbO$_{2}$ film was transferred under ambient conditions from the PLD camber to the LEED and STM/S chambers. For surface preparation, the sample was subjected to Ar$^{+}$ sputtering and annealing cycles, \textcolor{red}{with a voltage of 750 V, emission current of 11 $\mu$A, chamber pressure of 10$^{-5}$ mbar, for 10 minutes, followed by annealing at 700 $^{\circ}$C for 1 hour.} The maximum temperature, which was measured by the system’s K-type thermocouple, was up to 700 $^{\circ}$C for 2 h in UHV. This procedure was employed in order to remove contamination and any Nb$_{2}$O${_5}$ layer from the surface. A vacuum suitcase with a base pressure of 2$\times$10$^{-10}$ mbar was employed for transferring sample between STM/LEED and XPS chambers under UHV environment \cite{Zhussupbekov2021}.

\subsection{B. Computational Details}
\begin{figure*}[t!]
    \centering
    \includegraphics[width=1\textwidth]{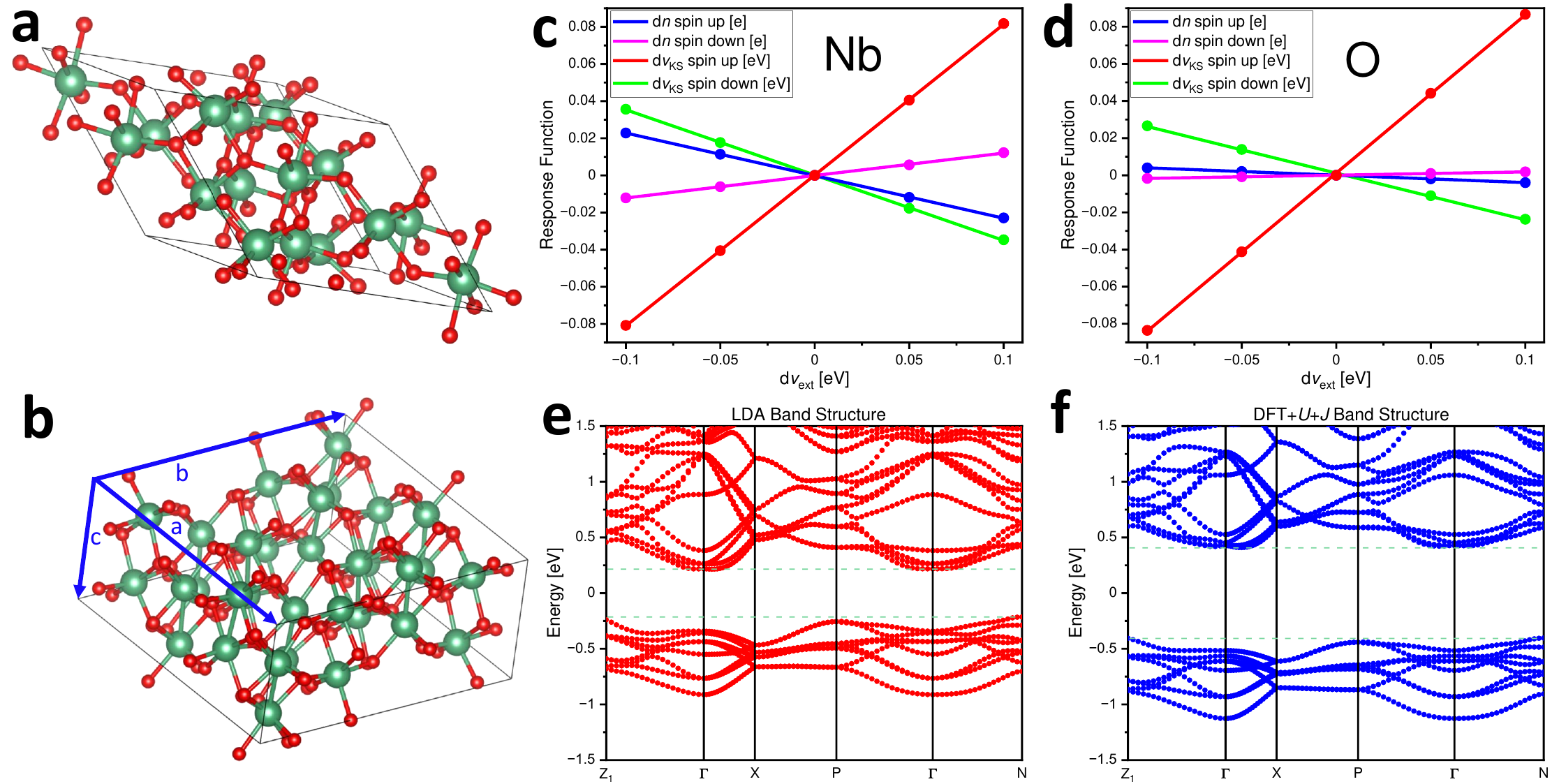}
    \caption{
     (a-b) Primitive and conventional units cells for \textcolor{red}{insulating} NbO$_2$ (Nb atoms shown in green, O in red), (c-d) Linear response plots for calculation of Hubbard $U$ and Hund's $J$. \textcolor{red}{(e-f) Band structure calculated by spectral function unfolding for LDA and DFT+$U$+$J$ respectively, using the implementation described in Refs. \cite{Constantinescu2015,Popescu2012} and recently demonstrated in Refs. \cite{Graham2021,Loh2021}.}
}
\label{fig:fig1}
\end{figure*}

The low temperature, \textcolor{red}{insulating} phase of NbO$_2$ takes on a BCT (distorted rutile) crystal structure. The primitive cell for this structure contains 48 atoms and is shown in Fig. \ref{fig:fig1}(a), alongside the 96 atom standard conventional unit cell in Fig. \ref{fig:fig1}(b). The crystallography of this material is most usually described in terms of the conventional cell, which is the cell we utilise in this study. It is worth noting that when discussing crystallographic directions some papers use the distorted rutile (\textrm{DR}) cell as a reference, and some use the almost equivalent rutile (\textrm{R}) cell. For clarity, the relations between some important directions are as follows: [100]$_{\textrm{DR}}$ $\simeq$ [110]$_\textrm{R}$, [110]$_{\textrm{DR}}$ $\simeq$ [100]$_\textrm{R}$. Crystal structures were visualised using the VESTA package \cite{Momma2011}.

Norm-conserving pseudopotentials generated with the OPIUM \cite{OPIUM} code were used throughout the study. DFT-LDA calculations were first carried out in the PWscf code of the Quantum Espresso suite \cite{Giannozzi2009,Giannozzi2017} to determine the cutoff energy and \textit{k}-point sampling necessary to converge the total energy to within $<$1 meV per atom, and the optimised ionic geometry/unit cell parameters. A cutoff energy of 60 Ry/815 eV and \textit{k}-point sampling of 2$\times$2$\times$4 for the standard conventional cell were deemed necessary. These parameters were then used to inform the parameters chosen for calculations in the ONETEP \cite{Skylaris2005,Prentice2020} linear-scaling DFT code, \textcolor{red}{which includes a DFT+$U$+$J$ implementation \cite{ORegan2012}}. A 2$\times$2$\times$4 supercell consisting of 1536 atoms was used with psinc spacing equal to \textit{a}/($3\cdot5\cdot7$)=0.4839 bohr along the distorted rutile \textit{a} axes and \textit{c}/(7$\cdot$13)=0.4934 bohr along the distorted rutile \textit{c} axis. A nonorthogonal generalised Wannier function (NGWF) cutoff radius of 10 bohr was utilised for both niobium and oxygen. For niobium, there were 10 NGWFs and 13 valence electrons per atom, compared to 4 NGWFs and 6 valance electrons per atom for oxygen. 

When using a supercell that is not simply a direct scaling of the primitive cell (i.e., when the transformation matrix $\mathbf{S}$ is not diagonal), it is important to verify that the high symmetry \textit{k}-points in the Brillouin zone are sampled within the supercell calculations. This is especially important for determining the band gap in an indirect gap \textcolor{red}{material} like BCT NbO$_2$. Following the method in Lloyd-Williams and Monserrat \cite{Lloyd2015}, in order for a \textit{k} point $\mathbf{q}$ to be commensurate with the supercell, the vector $\mathbf{Q}$ resulting from the transformation matrix $\mathbf{S}$ acting on $\mathbf{q}$ must contain only integers (Eq. 1.). \textcolor{red}{The necessary equations for determining if a high symmetry point is sampled are}

\begin{equation}
    \mathbf{S} \mathbf{q} = \mathbf{Q}
\end{equation} 

$$ \mathbf{S} \begin{pmatrix}0\\{}^1{\mskip -5mu/\mskip -3mu}_2\\0\end{pmatrix} = \begin{pmatrix}0\\1\\2\end{pmatrix} \quad \mathbf{S} \begin{pmatrix}{}^1{\mskip -5mu/\mskip -3mu}_2\\{}^1{\mskip -5mu/\mskip -3mu}_2\\-{}^1{\mskip -5mu/\mskip -3mu}_2\end{pmatrix} = \begin{pmatrix}2\\0\\0\end{pmatrix} \quad \mathbf{S} \begin{pmatrix}{}^1{\mskip -5mu/\mskip -3mu}_4\\{}^1{\mskip -5mu/\mskip -3mu}_4\\-{}^1{\mskip -5mu/\mskip -3mu}_4\end{pmatrix} = \begin{pmatrix}0\\0\\2\end{pmatrix} $$
~~~$\mathbf{q}=$ N~~~~~~~~~~~~~~~~~$\mathbf{q}=$ Z~~~~~~~~~~~~~~~~~~$\mathbf{q}=$ P

\vspace*{1\baselineskip}

In this case, we can see that the high symmetry points are correctly sampled, in particular the N point where the valence band maximum is located. 

In order to \textcolor{red}{better} account for the exchange and correlation effects present \textcolor{red}{in this system than is provided for in the underlying LDA functional}, we utilise the simplified, rotationally invariant DFT+$U$+$J$ method of the form presented in Ref. \cite{Himmetoglu2011}. \textcolor{red}{Functionals of this type are} widely used for calculations on metal oxides to account for the self interaction error (SIE) present in approximate exchange correlation functionals. \textcolor{red}{ The Hund's $J$ term, in particular, is associated with static correlation error, and the DFT+$U$+$J$ functional used here was designed to improve the description of spin-flip interactions. It has recently been tested, for example in Refs. \cite{Lim2016,Linscott2018,Orhan2020,Winczewski2022}, and some of its limitations have been explored in Ref. \cite{Burgess2023}}. \text

\begin{table}[b!]
\caption{\label{tab:table1}
Calculated Hubbard $U$ and Hund's $J$ parameters for the Nb 4\textit{d} and O 2\textit{p} subspaces in units of eV.
}
\begin{ruledtabular}
\begin{tabular}{l c @{\hspace{2\tabcolsep}} c}
\textrm{ }&
\textrm{Niobium}&
\multicolumn{1}{c}{Oxygen}\\
\colrule
$U$ & 2.48 & 9.02 \\
$J$ & 0.23 & 0.90 \\
$U_{\mathrm{eff}}=U-J$ & 2.25 & 8.12 \\ 
$U_{\mathrm{full}}=U-2J$ & 2.02 & 7.22 \\
\end{tabular}
\end{ruledtabular}
\end{table}
\textit{Ab-initio} calculations of the LDA-appropriate Hubbard $U$ and Hund's $J$ parameters were carried out within the minimum-tracking linear-response approach described in Linscott et al. \cite{Linscott2018}, and recently applied successfully to rutile and anatase TiO$_2$ \cite{Orhan2020}. For a closed shell system such as that of NbO$_2$, the `scaled 2$\times$2' approach reduces to the `simple 2$\times$2' approach as described in the aforementioned papers. \textcolor{red}{Within this method, $U$ and $J$ are determined by the matrix elements $f^{\sigma\sigma'}$  as in Eq. 2, with these matrix elements being determined by Eq. 3. The necessary equations are}
\begin{equation}
    U = \frac{1}{2} (f^{\uparrow \uparrow} + f^{\uparrow \downarrow}) \qquad J = -\frac{1}{2} (f^{\uparrow \uparrow} - f^{\uparrow \downarrow})
\end{equation}
\begin{equation}
    f^{\sigma\sigma'} = \left[ \left( \frac{\delta v_{\mathrm{KS}}}{\delta v_{\mathrm{ext}}} - \mathds{1}\right) \left(\frac{\delta n}{\delta v_{\mathrm{ext}}}\right)^{-1}  \right]^{\sigma\sigma'}
\end{equation}
\textcolor{red}{Based on this procedure we obtain $U$ and $J$ values for niobium and oxygen, as summarised in Table \ref{tab:table1}. The relatively large first-principles oxygen $U$ value and low metal $U$ value is consistent with literature findings, generated using other DFT codes, on diverse closed-shell early transition-metal oxides \cite{Kirchner2021,Lambert2023,Guy2022}. This trend reflects, among other factors, that the O 2\textit{p} orbitals are at least as localized, and sometimes considerably more so, than the metal d orbitals in these systems (see Fig. 3 of Ref. \cite{Lambert2023}). Interestingly, and similarly to the findings on TiO$_2$ in Ref. \cite{Orhan2020}, the computed $J$ values for both species with LDA are rather low in this
system, at approximately one-tenth of the corresponding $U$.}

\section{III. Results}
\subsection{A. Band Gap}

For the conventional 96 atom unit cell our LDA calculations give lattice parameters of \textit{a}=13.4445 \AA \, (2.15\% lower than experimental value 13.7020 \AA) and \textit{c}=5.9394 \AA \, (0.89\% lower than the experimental value of 5.9850 \AA) \cite{Bolzan1994,Dhamdhere2016}. \textcolor{red}{The most crucial aspect of the NbO$_2$ crystal structure is the Nb-Nb dimerization along the \textit{c}-axis ([001] direction). The lattice parameter we obtain along this direction is very close to the experimental value, which is vital since the electronic structure of NbO$_2$ is known to be highly sensitive to the dimerization \cite{Fajardo2021,OHara2015}. The short and long Nb-Nb bond lengths at room temperature are 2.71 \AA \ and 3.30 \AA \ respectively \cite{Bolzan1994,Pynn1976,Pynn1976_1} compared to the values obtained in our calculation of 2.66 \AA \ and 3.31 \AA, again close to the room temperature values. Of course it should be noted that the bond length varies as a function of temperature (roughly 0.1 \AA \ between 100 $^{\circ}$C and 800 $^{\circ}$C \cite{Fajardo2021,Wahila2019}), but no low temperature measurements of the bond length exist in the literature to the authors' knowledge.}

Our LDA calculated band gap of 0.39 eV is slightly higher than other LDA calculations \cite{OHara2014} (possibly due to details of the pseudopotential construction), but is still underestimating the true gap by at least a factor of 2. A similar trend emerges for the other spectral features, with slight \textcolor{red}{increase} over previous LDA calculations yet still not meaningfully approaching the experimental values. \textcolor{red}{The LDA calculated band structure is shown in Fig. \ref{fig:fig1}(e), with the valance band maximum (VBM) at the N point, and the conduction band minimum (CBM) at the $\Gamma$ point in line with previous reports \cite{OHara2014,Kulmus2021}.}

\begin{table}
\caption{\label{tab:table2} Summary of experimental and theoretical values for the band gap. Theoretical values pertain to that of bulk NbO$_2$.} 
\begin{ruledtabular}
 \begin{tabular}{||c c||} 
 \hline
 Method & Band Gap [eV] \\ [0.5ex] 
 \hline\hline
 LDA \cite{OHara2014} & 0.35 \\ 
 \hline
 SCAN \cite{Fajardo2021} & 0.48 \\
 \hline
 HSE \cite{OHara2014} & 1.48 \\
 \hline
 DFT+cDMFT \cite{Brito2017} & 0.73 \\
 \hline
 LDA \cite{Wahila2019} & 0.058 \\
 \hline
 PBE \cite{Wahila2019} & 0.127 \\
 \hline
 mBJ \cite{Wahila2019} & 0.844 \\
 \hline
 GW-BSE \cite{Kulmus2021} & 0.98 \\
 \hline
  &  \\
 \hline
 LDA (this work) & 0.39 \\
 \hline
 DFT+$U^d$ (this work) & 0.83 \\
 \hline
 DFT+$U^d_{\mathrm{eff}}$ (this work) & 0.78 \\
 \hline
 DFT+$U^d$+$J^d$ (this work) & 0.74 \\
 \hline
 DFT+$U^{d,p}$ (this work) & 0.92 \\
 \hline
 DFT+$U^{d,p}_{\mathrm{eff}}$ (this work) & 0.86 \\
 \hline
 DFT+$U^{d,p}$+$J^{d,p}$ (this work) & 0.80 \\
 \hline
  &  \\
 \hline
 Resistance vs Temperature \cite{Sakai1985} & 1.23 \\
 \hline
 Resistance vs Temperature \cite{Stoever2020} & 0.88 \\
 \hline
 Scanning Tunnelling Spectroscopy & 1.05 $\pm$ 0.29\\ 
 (this work) & \\
 \hline
 Inverse Photoemission \cite{Posadas2014} (4nm film) & 1.0 (lower bound) \\ 
 \hline
 Optical Absorbtion \cite{Stoever2020} (T=4K) & 0.85 \\ 
 \hline
 Film Absorption Edge \cite{Lee1984} & 0.88 \\ 
 \hline
 Optical Ellipsometry \cite{OHara2014} (indirect gap) & 0.7 \\ 
 \hline
\end{tabular}
\end{ruledtabular}
\end{table}

The linear response plots used to calculate the $U$ and $J$ for the Nb and O are shown in Fig. \ref{fig:fig1}(c-d), and reflect good linear response. There are a number of different ways of implementing the Hubbard $U$ and Hund's $J$ corrections. In many published works, the correction is applied only the transition metal \textit{d} orbitals (such as in Ref. \cite{OHara2014}). We consider three possibilities for the specific form of the correction. The first option is to only add the $+U$, \textcolor{red}{not including explicit opposite-spin interaction terms in the corrective functional} (effectively treating $J$ as being zero). Alternatively, the Dudarev functional \cite{Dudarev1998} additionally accounts for exchange-related corrections arising from interactions between like spins, by taking $U_{\textrm{eff}}=U-J$. Finally, corrections arising from interactions between both like and unlike spins \textcolor{red}{can be obtained by carrying out full spin polarised calculations with DFT+$U$+$J$, or equivalently for non spin-polarised systems, as presented in Ref. \cite{Orhan2020} by using the Dudarev functional with $U_{\textrm{full}}=U-2J$ and $\alpha=J/2$}. In all three of these cases, we can then also apply the same form of the correction to the O 2\textit{p} subspace, resulting in a total of 6 possible implementations of these $U$ and $J$ values.

Table \ref{tab:table2} shows the calculated band gaps for the 6 different corrections tested, along with experimental and theoretical values from the literature. Interestingly we can see that even though the band gap exists between Nb 4\textit{d} states, adding the correction to the O 2\textit{p} subspace does have a small effect on the calculated gap. The gap increased by about 0.06-0.09 eV or roughly 10\% across the three different forms of the functional. \textcolor{red}{The DFT+$U^{d,p}$+$J^{d,p}$ calculated band structure is shown in Fig. \ref{fig:fig1}(f). We see that the VBM remains at the N point, however the CBM is shifted away from the $\Gamma$ point along the $\Gamma$-X direction. This is in line with Ref. \cite{OHara2014}, where they observed a similar change in the band curvature resulting from applying a Hubbard $U$ correction.}

\textcolor{red}{In the supplemental information \cite{SI1} (see also references \cite{Huang2011,Wong2014} therein) we detail the structural and chemical characterisation of the thin film. Crucially, we see no evidence of contamination from Nb$_2$O$_5$ from core level XPS (Fig. SI1), as has been present in some previous studies \cite{Fajardo2021,Wahila2019}.}

\begin{figure*}[t!]
    \centering
    \includegraphics[width=1\textwidth]{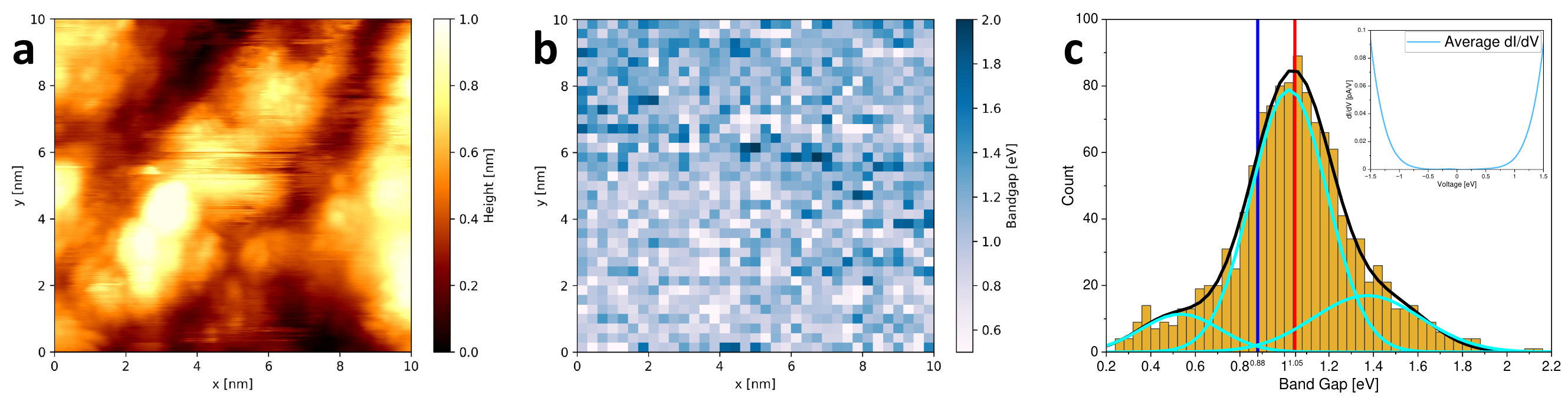}
    \caption{
     (a) Constant current STM image of the (100) oriented film ($V$\,=\,-2.5\,V and $I$\,=\,70\,pA), (b) Band gap map from same area consisting of 35\,$\times$\,35=1225 $I(V)$ curves, (c) statistical analysis of the 1225 measured band gaps (red line represents mean of (1.05 $\pm$ 0.29) eV, blue line represents electrical band gap measurement from ref. \cite{Stoever2020} of 0.88 eV), black line represents trimodal gaussian fit with distributions centred around (0.53, 1.03, 1.37) eV (individual distributions shown in light blue), inset shows the average $\mathrm{d}I/\mathrm{d}V$ curve.
}  
\label{fig:fig3}
\end{figure*}

\textcolor{red}{STM of this film shows a relatively smooth surface with an RMS roughness of $\simeq$ 2 \AA\ over 10 nm$^{2}$ (Fig. \ref{fig:fig3}(a)), with a granular structure, often observed for oxide thin films \cite{Kurnosikov2001,Stavale2012}. Due to the high reactivity of Nb with oxygen, preparing a smooth enough surface for obtaining atomic resolution is likely not possible with a thin film (as compared with the less reactive Ti, for which atomic resolution STM images can be obtained on thin films \cite{Ashworth2001,Shi2020}. Much more aggressive sputter anneal cycles would be necessary, which could only be performed on a single crystal.} Grid spectroscopy measurements were performed on the same area as the STM image in Fig. \ref{fig:fig3}(a). Fig. \ref{fig:fig3}(b) shows the resulting band gap map from this grid measurement. 35\,$\times$\,35 individual $I(V)$ measurements were performed on the grid. The CCM scanning parameters utilised to move tip between points were $V$\,=\,-2.5\,V and $I$\,=\,70\,pA. The voltage has been swept between $\pm$\,2\,V. Fig. \ref{fig:fig3}(c) shows the statistical analysis of the grid STS. The inset of Fig. \ref{fig:fig3}(c) illustrate averaged dI/dV spectrum of the whole grid. These STS measurements demonstrate that the majority of the measured local band gaps vary from roughly 0.75 eV to 1.35 eV, with a statistical mean of (1.05 $\pm$ 0.29) eV. We can fit this with three Gaussian distributions as shown in Fig. \ref{fig:fig3}(c). There are a number of possible effects that could contribute to the width of the distribution. Excess oxygen could cause an increase in the local surface band gap as measured by STS, and we do indeed observe a slightly lower Nb:O ratio of 31\%:69\% from \textit{in-situ} XPS measurements. Therefore, we suggest that the areas of higher band gap likely arise due to an excess of oxygen on and near the surface. There are a number of ways this excess oxygen could be incorporated at the surface. Some reports of thin film growth show both Nb$^{4+}$ and Nb$^{5+}$ present at the surface \cite{Posadas2014,Fajardo2021}, indicating the presence of some higher oxides in parts of the surface \cite{Zhussupbekov2020}. However, for our surface as scanned, core level XPS on the Nb 3\textit{d} state shows only an Nb$^{4+}$ component, indicating no higher valence oxide layer on the surface. Analysis of the oxygen 1\textit{s} core level state shows two chemical states of oxygen. \textcolor{red}{Further sputter anneal cycles reduce the intensity of the secondary oxygen peak (see SI Fig. 1(d-e)), suggesting that this oxygen is more weakly bound to the surface. Therefore we believe that this component likely comes from oxygen adatoms/hydroxide groups at the surface, which have been observed on other rutile (110) systems such as TiO$_2$ \cite{Epling1998, Zhang2019, Skelton2007, Wen2020}.} LEED also supports this viewpoint (Fig. SI1(a)), as we see no evidence of a major surface reconstruction caused by the formation of additional oxides. Instead, LEED measurements confirm the long range crystalline order with the expected distorted rutile (100)/rutile (110) like symmetry.

\subsection{B. Band widths}

\begin{figure*}[t!]
    \centering
    \includegraphics[width=0.87\textwidth]{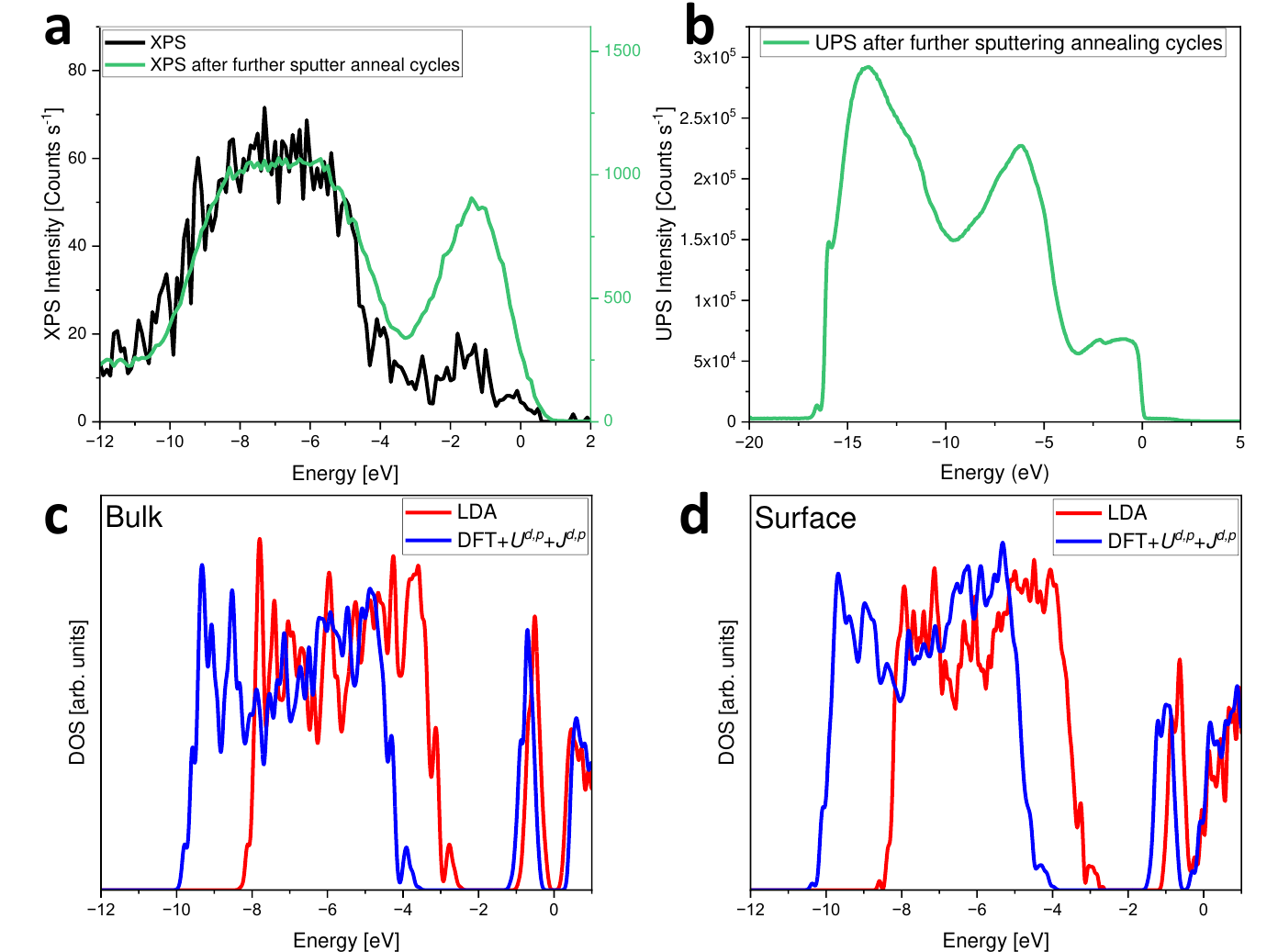}
    \caption{
     (a) Measured valence band XPS for the NbO$_2$(100) film before and after further sputter anneal cycles, (b) UPS measured after further sputter anneal cycles, (c) LDA and DFT+$U$+$J$ calculated DOS for the bulk, (d) LDA and DFT+$U$+$J$ calculated DOS for the NbO$_2$(100) slab.
}  
\label{fig:fig4}
\end{figure*}

\textcolor{red}{The initially measured valence band XPS (Fig. \ref{fig:fig4}(a), black curve) shows characteristics similar to that observed in Ref. \cite{Posadas2014} with low counts per second, and a diminished intensity of the Nb 4$d$ peak. We believe that this is due to oxygen contamination at the surface, likely in the form of oxygen adatoms due to the lack of any other Nb components observed in core-level XPS. Further sputter anneal cycles on our film dramatically improve the counts per second, and we see an increase in the intensity of the Nb 4$d$ peak (accompanied by a decrease in the secondary oxygen component in core-level XPS). This type of spectra (Fig. \ref{fig:fig4}(a), green curve) is consistent with that observed in Refs. \cite{Wahila2019,Wong2014,OHara2014}. Across both of the observed spectra we see values for the band widths and $p-d$ gap that are generally consistent with range of values from previous reports (see Table \ref{tab:table3}). While in some works a sharp $p-d$ gap is observed, we do not see this in our spectra. This could be due to weak disorder and/or the remaining oxygen contamination not removed by sputter anneal cycles. Ultra-violet photoelectron spectroscopy (UPS) measurements (Fig. \ref{fig:fig4}(b)) show that the surface exhibits metallic behaviour.}

Since XPS is a surface sensitive technique, it is possible that modifications of the electronic structure at the surface could contribute to the lack of agreement between the experiment and theory. To this end, in addition to the bulk calculations, we also consider a (100)$_{\textrm{DR}}$ surface, using a slab with 8 NbO$_2$ layers (24 atomic layers, symmetric to ensure no spurious dipole moment). Fig. \ref{fig:fig4}(c-d) shows the calculated DOS for the bulk and surface respectively, for LDA and O 2\textit{p}. Table \ref{tab:table3} shows the calculated band widths and $p$-$d$ gaps for our calculations, along with experimental and theoretical values from the literature. 

For the bulk calculations, we see that when $U$ and $J$ corrections are applied to both the Nb 4$d$ and O 2$p$ subspaces, we get values for the $p$ bandwidth that are comparable with experiment. The $p$-$d$ gap is also improved when applying the correction to both atoms, though overshooting the experimental values. This reinforces the findings of Ref. \cite{Orhan2020}, namely that for a charge-transfer gap like this $p$-$d$ sub gap, appropriate corrections for both subspaces must be included to obtain the correct value for that gap. For the 4$d$ bandwidth, there is marginal improvement when including the $U$ and $J$ correction, but still well shy of the experimentally measured values.

Moving to the slab calculations, we see that there is a clear non-zero DOS at the Fermi level. \textcolor{red}{This leads to an increase the Nb 4\textit{d}$_{xy}$ band width}, and as a result slightly decreases the \textit{p}-\textit{d} gap. Unsurprisingly, the O 2\textit{p} band width is largely unaffected by this effect. We \textcolor{red}{see} that overall, the DFT+$U^{d,p}$+$J^{d,p}$ slab calculation presents solid agreement with the range of experimentally measured values. The Nb 4\textit{d}$_{xy}$ width is increased to 1.5 eV, in line with experiment. The \textit{p}-\textit{d} gap is now 2.2 eV, much closer to the experimental value. The O 2\textit{p} width is unchanged from the bulk calculation of the same functional, remaining in good agreement with experiment. It therefore seems that the underestimation of the calculated bulk Nb 4\textit{d}$_{xy}$ band width compared to the measured value is not purely \textcolor{red}{related to the} to the exchange-correlation functional, but also due to metallisation at the surface.

\begin{table*}[t!] 
\caption{\label{tab:table3} Summary of experimental and theoretical values for the valence band XPS spectral features. Results from the most elaborate theoretical treatment, \textcolor{red}{including both surface effects, as well as Hubbard $U$ and Hund $J$ corrections, are highlighted in bold.}}
\begin{ruledtabular}
 \begin{tabular}{||c c c c||}
 \hline
 Method & \textit{p-d} Gap [eV] & \textit{p} bandwidth [eV] & \textit{d} bandwidth [eV] \\ [0.5ex] 
 \hline\hline
 LDA \cite{OHara2014} & 1.58 & 5.55 & 0.70 \\ 
 \hline
 HSE \cite{OHara2014} & 1.79 & 5.76 & 0.71 \\
 \hline
 GW-BSE \cite{Kulmus2021} & 2.4 & 5.7 & 0.79 \\
 \hline
  & & & \\
 \hline
  LDA bulk (this work) & 1.51 & 5.65 & 0.80 \\
 \hline
  DFT+$U^d$ bulk (this work) & 1.16 & 5.70 & 0.91\\
 \hline
 DFT+$U^d_{\mathrm{eff}}$ bulk (this work) & 1.17 & 5.70 & 0.88 \\
 \hline
 DFT+$U^{d}$+$J^{d}$ bulk (this work) & 1.20 & 5.71 & 0.86 \\
 \hline
 DFT+$U^{d,p}$ bulk (this work) & 2.80 & 6.90 & 0.87\\
 \hline
 DFT+$U^{d,p}_{\mathrm{eff}}$ bulk (this work) & 2.64 & 6.65 & 0.88 \\
 \hline
  DFT+$U^{d,p}$+$J^{d,p}$ bulk (this work) & 2.31 & 6.40 & 0.85\\
 \hline
  LDA (100) slab (this work) & 1.48 & 5.65 & 0.95 \\
 \hline 
  \textbf{DFT+}$\boldsymbol{U^{d,p}}$\textbf{+}$\boldsymbol{J^{d,p}}$ \textbf{(100) slab (this work)}  & \textbf{2.20} & \textbf{6.40} & \textbf{1.50} \\
 \hline
  & & & \\
 \hline
 XPS (this work, initial) & $\approx$ 2.0 & $\approx$ 6.2 & $\approx$ 1.5 \\
 \hline
 XPS (this work, after further sputter anneal cycles) & $\approx$ 1.7 & $\approx$ 6.4 & $\approx$ 1.9 \\
 \hline
 XPS \cite{Posadas2014} & $\approx$ 1.8 & $\approx$ 6.3 & $\approx$ 1.5 \\ 
 \hline
 XPS \cite{OHara2014} & $\approx$ 2.0 & $\approx$ 6.2 & $\approx$ 1.7 \\ 
 \hline
  XPS \cite{Wong2014} & $\approx$ 2.0 & $\approx$ 6.8 & $\approx$ 1.2 \\ 
 \hline
  HAXPES \cite{Wahila2019} & $\approx$ 1.9 & $\approx$ 6.0 & $\approx$ 1.5 \\ 
 \hline
\end{tabular}
\end{ruledtabular}
\end{table*}

\begin{figure*}[t!]
    \centering
    \includegraphics[width=1\textwidth]{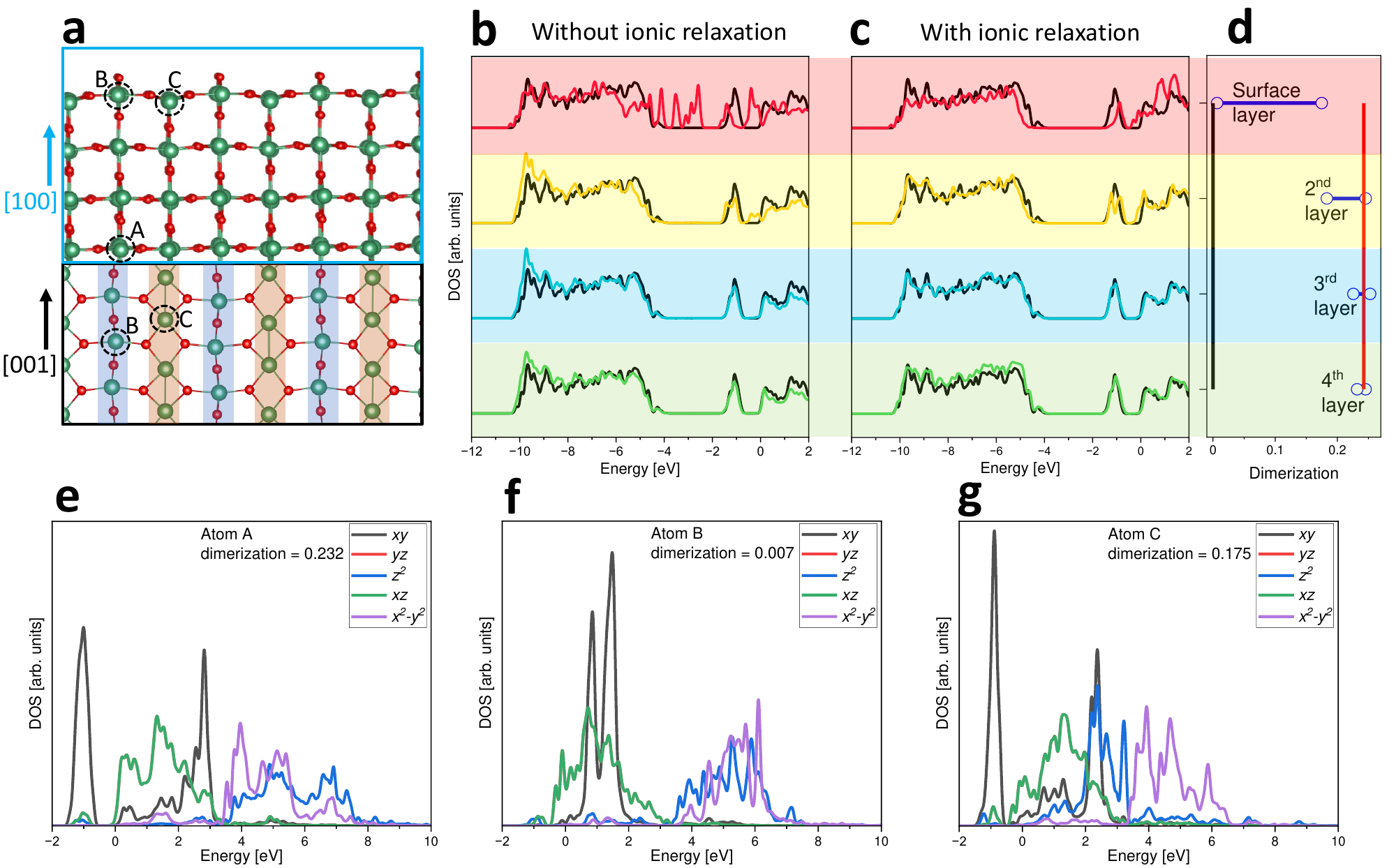}
    \caption{
     (a) Surface crystal structure showing side view (blue panel) and top down view (black panel) (b-c) Layer resolved local density of states for the (100) slab with shifted bulk DOS superimposed without and with ionic relaxation (d) dimerization (=$\frac{\textrm{long bond}}{\textrm{short bond}}-1)$ versus layer number (e-g) $m_l$ projected DOS for bulk atom, and the two types of surface atom. 
     }
\label{fig:fig5}
\end{figure*}

\textcolor{red}{The optimised geometry of the slab shows that the dimerization collapses almost entirely for the Nb atoms under the bridging oxygens (blue highlighted rows in Fig. \ref{fig:fig5}(a)). The dimerization is also reduced relative to the bulk for the 5-fold Nb atoms and those in the 2$^{\textrm{nd}}$ layer of the slab. Fig. \ref{fig:fig5}(d) shows the dimerization plotted as a function of depth into the surface. It may be tempting then to ascribe the metallisation of the surface to this breakdown in dimerization, however there is another competing effect which can change the surface band structure, namely the reduced Coulomb repulsion at the surface. Fig. \ref{fig:fig5}(b-c) show the layer resolved DOS for the slab without and with ionic relaxation. We see that even when the ionic geometry is kept fixed (no breakdown in dimerization allowed) metallicity still emerges at the surface.}

\textcolor{red}{In order to further quantify these two effects, we plot the DOS, decomposed by magnetic quantum number $m_l$ for the Nb atoms in the slab. We utilise the method developed in Ref. \cite{Aarons2019}, and use pseudo-atomic orbitals as projectors. The coordinate system chosen for the basis functions is such that the dimerization lies in the $xy$ plane. We first note that for the bulk Nb atoms (Fig. \ref{fig:fig5}(e) in the slab our m$_l$ projected DOS resembles that of Ref. \cite{Kulmus2021}, with the $d_{xy}$/$d_{xy}^*$ splitting caused by dimerization, with degenerate $d_{yz}$ and $d_{xz}$ orbitals. Now looking at the plots for the surface atoms, we see that for the Nb atoms under the bridging oxygens (Fig. \ref{fig:fig5}(f)), (dimerization $\simeq$ 0) the $d_{xy}$/$d_{xy}^*$ splitting has disappeared, as is the expected behaviour from a Peierls type splitting. We also see that the energy of the $d_{yz}$ and $d_{xz}$ orbitals has decreased, as expected due to the reduction in Coulomb repulsion at the surface. For the 5-fold surface Nb atoms (Fig. \ref{fig:fig5}(g)), we see the same reduction in energy of $d_{yz}$ and $d_{xz}$ orbitals, but with the $d_{xy}$/$d_{xy}^*$ splitting still present. Looking at both plots, we notice that the non-zero DOS at the Fermi level is coming mostly from the $d_{yz}$ and $d_{xz}$ orbitals, indicating that the lowering in energy of these orbitals is the underlying cause of the metallicity at the surface. Finally, we note some of the outstanding discrepancies between experiment and theory in the \textit{p}-\textit{d} gap region, due to neglected effects in the latter. In particular, referring to the XPS data in Fig. \ref{fig:fig4}(a-b), there is non-zero spectral weight between -3 and -1 eV, where in the calculated (generalized Kohn-Sham) spectra there is a clear \textit{p}-\textit{d} gap. There are certainly surface disorder and finite-temperature effects that are neglected in the simulations. Referring to Fig. \ref{fig:fig5}(b), it is clear that surface ionic displacements can readily populate this region with spectral features. Furthermore, there are many-electron effects that cannot be seen at the level of theory used here. These include plasmon satellite bands, Hubbard bands, and enhanced spectral splitting due to strong and orientation-dependent intra-orbital interactions associated exchange and correlation effects. Noting the significant first-principles Hubbard U for O 2\textit{p} orbitals, exceeding their band-width, we cannot rule out the possibility of unconventional strong 2\textit{p} electron correlation effects, as well as the more usually anticipated Nb 4\textit{d} ones.}

\section{IV. Conclusions}
Overall, we find that the $U$ and $J$ values for both Nb 4\textit{d} and O 2\textit{p} orbitals in NbO$_2$ can be straightforwardly calculated due to well behaved linear response. The inclusion of a correction based on these values can significantly improve agreement with experimentally measured spectral quantities, demonstrating the efficiency of this first-principles method in capturing the essential physics of the material. \textcolor{red}{The band gap and band widths are relatively robust against the exact form of the correction used, due in part to the relatively small Hund's $J$ value for both Nb 4\textit{d} and O 2\textit{p}. At approximately one-tenth of the corresponding Hubbard $U$ values, the low Hund's $J$ values indicate an already good treatment of static correlation effects at the LDA level in this system.} 

Additionally, in order to explain the measured XPS spectra we must account for surface effects, \textcolor{red}{due to the emergence of metallic behaviour. It seems that although the dimerization does break down at the surface, it is not the driving force behind the transition from insulating to metallic behaviour at the surface. Therefore it cannot be considered a purely Peierls transition. Our results show that a transition from insulating to metallic behaviour can be induced in NbO$_2$ from a change in the Coulomb potential \textcolor{red}{at the surface. This is the case irrespective of whether corrections for correlation effects at the DFT+$U$+$J$ level are incorporated. Indeed, due to the relatively small Hund's $J$ parameters, it is sufficient to include corrections beyond LDA at the Hubbard $U$ level alone in order to achieve good agreement with spectroscopic measurements. Such corrections do not explicitly address correlation within this formalism, in the strict sense of the word, but only Coulomb repulsion and exchange. Thus, our results give only partial credence to the reports that electronic correlation play an important role in this system \cite{Rana2019,Brito2017}, and we cannot rule out agreement with other reports which claim correlations to be unimportant \cite{Wahila2019,Kulmus2021}}.}

\textcolor{red}{We emphasize, however, that our findings pertain primarily to surface effects, as distinct from bulk effects such as the temperature dependent metal-insulator transition. We have not, for example, explored the relationship between bulk dimerization or finite temperature effects with the calculated $U$ and $J$ parameters, which remains an avenue for future research.} Our measurements also show the presence of excess oxygen on and near the surface, which could also play a role in modifying the observed spectra. However, since we only see evidence of oxygen adatoms and not a continuous layer of a higher valence oxide, we do not expect that these adatoms could account for the change we predict between the surface and bulk properties (e.g., the 0.5 eV widening of the Nb \textit{d}$_{xy}$ band width observed in Fig. \ref{fig:fig4}(c-d)).

In order to further confirm that the metallicity is confined to the surface, the surface and bulk contributions to the valence band photoelectron signal would need to be decoupled, using a technique sensitive to both the bulk and surface. Hard X-ray photoelectron Spectroscopy measurements were reported in Ref. \cite{Wahila2019}, but only for a single photon energy. Multiple measurements varying the photon energy would be needed to decouple the surface and bulk contributions, and to determine whether or not the observed Nb \textit{d}$_{xy}$ band width comes mostly from the surface effect calculated here, or a combination of bulk and surface. \textcolor{red}{Angle resolved XPS/UPS could also shed light on this effect.}

This metallisation at the surface could give rise to new applications and new physics. Anisotropic 2D electron gases have attracted significant attention recent years \textcolor{red}{due to their novel properties such as 2D superconductivity, unique magnetic phases, and nonlinear transport characteristics \cite{Annadi2013, Lev2018, Lee2012, Wang2014}}. With the top two layers near the surface becoming metallic, while the bulk remains insulating, a 2D electron gas may form. Additionally, NbO$_2$ is known to have significant anisotropy in its conductivity, with the lowest resistance along the dimerization axis \cite{Sakai1985,Belanger1974,Stoever2020}. Due to the partial breakdown of dimerization at the surface, we would expect the anisotropy of the conductivity to be even higher for these metallic surface electrons, giving rise to a highly anisotropic 2D electron gas.

\section{Acknowledgements}
This work was supported by the Irish Research Council (IRC) Laureate Award (IRCLA/2019/171), Science Foundation Ireland. The authors wish to acknowledge the Irish Centre for High-End Computing (ICHEC) for the provision of computational facilities and support, as well as the Trinity Centre for High Performance Computing (Boyle cluster). K.Z would like to acknowledge support from the Ministry of Science and Higher Education of the Republic of Kazakhstan (No. AP19174589). \textcolor{red}{This work was further supported (D.D.O'R) by Science Foundation Ireland (No. 19/EPSRC/3605) and the Engineering and Physical Sciences Research Council (No. EP/S030263/1).} A.Z. and K.Z. would also like to acknowledge funding from IRC through No. GOIPD/2022/443 and No. GOIPD/2022/774 awards. J.E.B. and J.S. acknowledge support from the Leibniz Competition project with the title ``Physics and Control of Defects in Oxide Films for Adaptive Electronics''. \textcolor{red}{D.D.O'R thanks N.D.M Hine and E.B. Linscott for helpful discussions.} We acknowledge and thank the ONETEP Developers Group.

\section{Bibliography}

\newpage

\title{\huge{Supporting Information}\\[5mm] \Large 
Reconciling the theoretical and experimental electronic structure of NbO$_2$
}

\maketitle

\renewcommand\thefigure{SI\arabic{figure}} 
\setcounter{figure}{0} 

\onecolumngrid

\begin{figure*}
    \centering
    \includegraphics[width=0.8\textwidth]{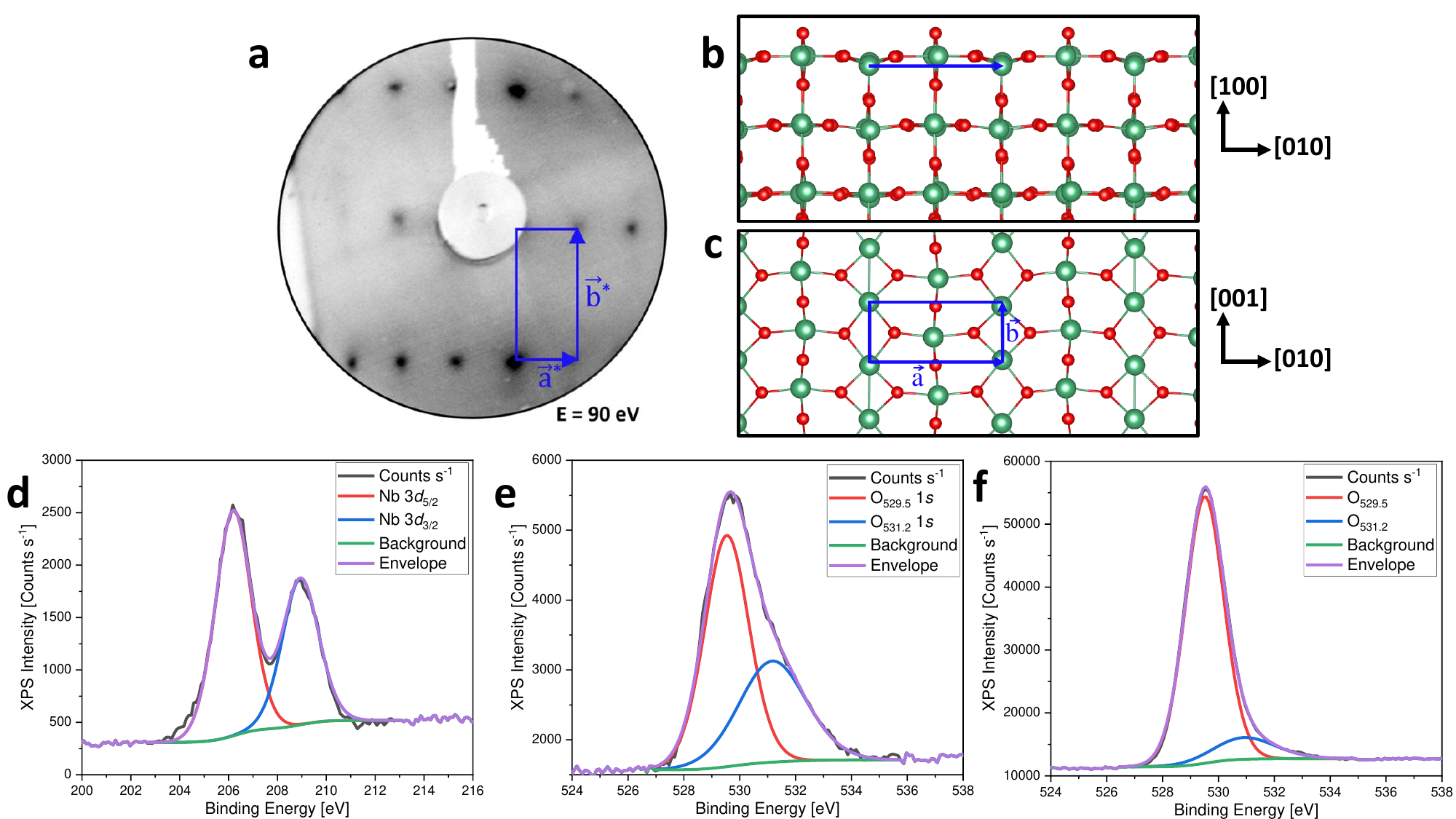}
    \caption{
     (a) Measured LEED pattern for NbO$_2$(100) film at E=90eV. (b-c) Side and top view of fully relaxed crystal structure. (d-e) Core level XPS of Nb 3\textit{d} and O 1\textit{s} levels respectively. 
}  
\label{fig:fig2}
\end{figure*}

The NbO$_{2}$ film was introduced to the UHV chamber and went through Ar ion sputtering and annealing cycles in a UHV environment at \SI{700}{\celsius} for 2\,h. The LEED measurement of the synthesised NbO$_{2}$ film reveals a long-range crystalline order as presented in Fig.\ref{fig:fig2}(a). The LEED pattern in Fig.\ref{fig:fig2}(a) was obtained with a primary beam energy of 90 eV, at room temperature. It shows a number of fairly sharp and well-resolved diffraction reflexes with two-fold symmetry. The directions of the reciprocal $\overrightarrow{a}^{*}$  and $\overrightarrow{b}^{*}$ vectors are indicated with blue colour on top of the LEED pattern \cite{Huang2011}.

Fig.\ref{fig:fig2}(b) shows the side and top views of the relaxed crystal geometry of the NbO$_{2}$ surface. The top-view of the surface is displaying the (100) plane of the NbO$_{2}$. The real-space vectors of the NbO$_{2}$ ($\overrightarrow{a}$ and $\overrightarrow{b}$) are shown in the schematic crystal structure model in blue colour for a better illustration. 

Fig.\ref{fig:fig2}(c-e) present the core level XPS of Nb 3\textit{d} and O 1\textit{s} levels, respectively. We can see for the Niobium core level in (c) that only one component corresponding to Nb$^{4+}$ is present \cite{Wong2014}. The lack of any Nb$^{5+}$ contribution (as has been reported in some other studies \cite{OHara2014,Fajardo2021,Wahila2019}) confirms that we are probing solely the NbO$_2$ surface structure. The initially measured oxygen core level shows two components (d), with the secondary component decreasing after further sputter anneal cycles (e). This suggests that the secondary component is likely arising from hydroxide groups or oxygen adatoms weakly bound to the surface.

\begin{figure*}
    \centering
    \includegraphics[width=0.85\textwidth]{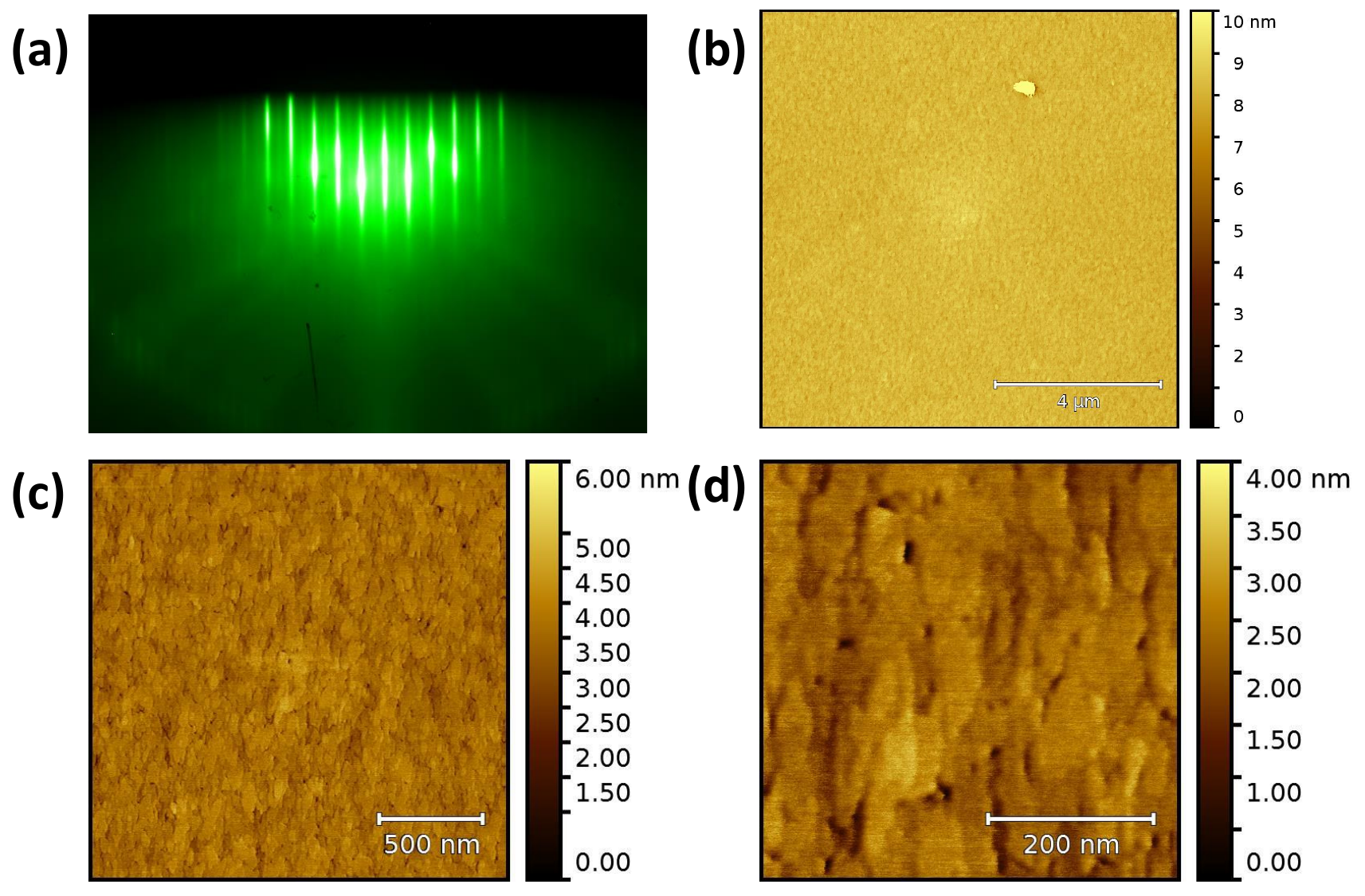}
    \caption{
     (a) RHEED image after deposition, (b-d) AFM images of (100) film.
}  
\label{fig:SI2}
\end{figure*}

\begin{figure*}
    \centering
    \includegraphics[width=0.85\textwidth]{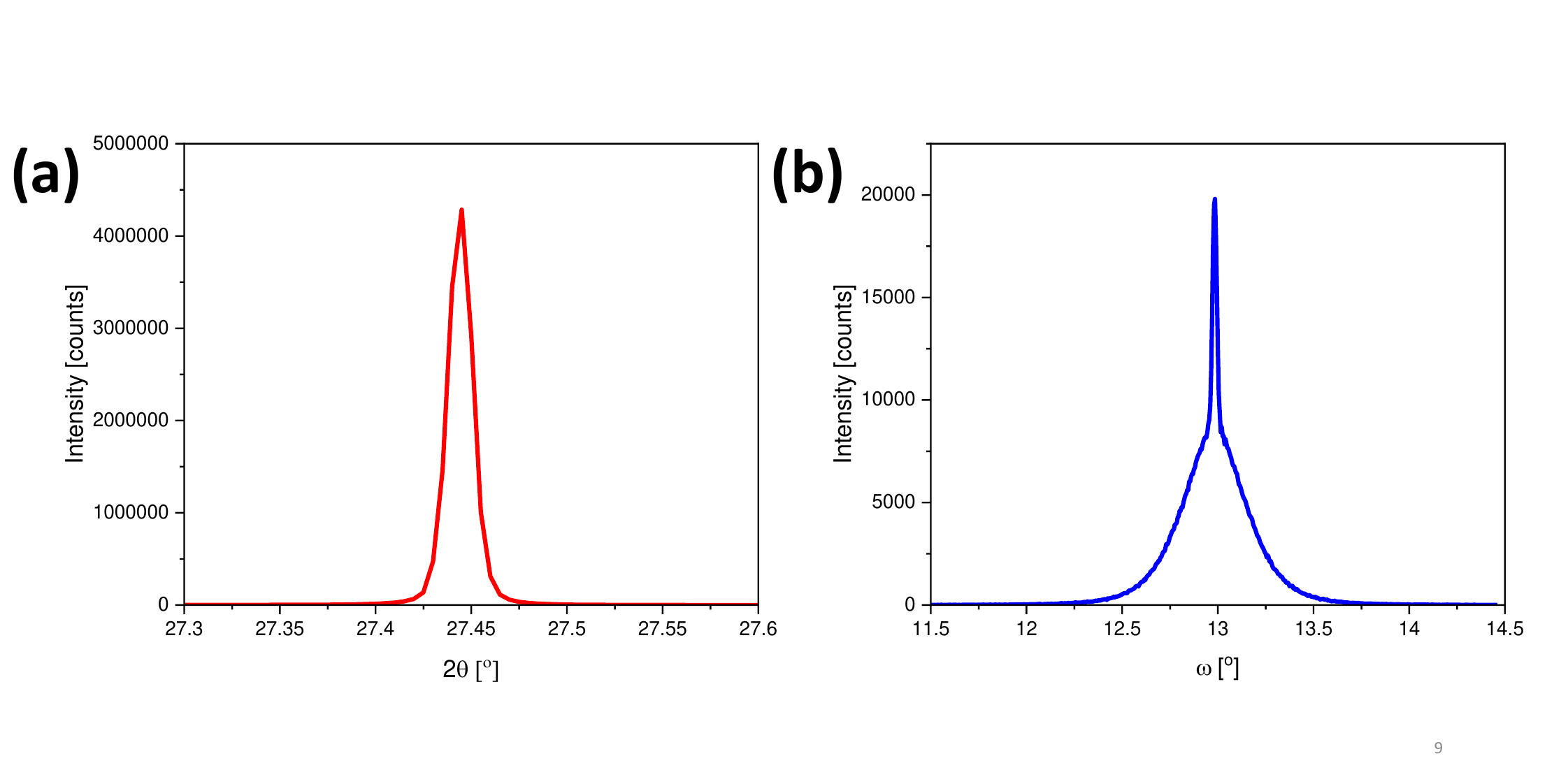}
    \caption{
     (a) $\theta-2\theta$ scan, (b) rocking curve.
}  
\label{fig:SI3}
\end{figure*}

Fig.\ref{fig:SI2} shows the RHEED and AFM images of the (100) film. The sharp reflexes and streak pattern demonstrates the high crystalline quality of the film, while the AFM images show that the film displays low roughness. Fig.\ref{fig:SI3} shows the XRD $\theta-2\theta$ scan and rocking curve, demonstrating the high quality single crystalline nature of the film. All in all, these measurements of RHEED, AFM, LEED, and XRD confirm the high crystalline quality of the epitaxial NbO$_{2}$ film, with a (100) distorted rutile (or (110) rutile) surface orientation.

\end{document}